\documentclass[aps,twocolumn, floats,preprintnumbers,showpacs,nofootinbib,prl]{revtex4}
\usepackage{graphicx}
\usepackage{url}
\usepackage{color}
\usepackage{amsmath}
\usepackage{amsfonts}
\usepackage{amssymb}
\def\beq{\begin{equation}}
\def\eeq{\end{equation}}
\def\beqa{\begin{eqnarray}}
\def\eeqa{\end{eqnarray}}

\newcommand{\bi}{\begin{itemize}}
\newcommand{\ei}{\end{itemize}}
\newcommand{\be}{\begin{equation}}
\newcommand{\ee}{\end{equation}}
\newcommand{\bea}{\begin{eqnarray}}
\newcommand{\eea}{\end{eqnarray}}
\newcommand{\nn}{\nonumber}

\def\ltap{\ \raise.3ex\hbox{$<$\kern-.75em\lower1ex\hbox{$\sim$}}\ }
\def\gtap{\ \raise.3ex\hbox{$>$\kern-.75em\lower1ex\hbox{$\sim$}}\ }

\begin{document}

\title{
2HDM Portal Dark Matter: LHC data and the Fermi-LAT 135 GeV Line
}

\author{Yang Bai$^{a,b}$, Vernon Barger$^{a}$, Lisa L.~Everett$^{a}$ and Gabe Shaughnessy$^{a}$
\\
\vspace{2mm}
${}^{a}$Department of Physics, University of Wisconsin, Madison, WI 53706, USA \\
${}^{b}$SLAC National Accelerator Laboratory, 2575 Sand Hill Road, Menlo Park, CA 94025, USA 
}

\pacs{12.60.Fr, 14.80.Ec}

\begin{abstract}
We study a two Higgs doublet model augmented by a scalar dark matter  particle that provides an excellent fit to the LHC Higgs data and the  Fermi-LAT 135 GeV line.  
The heavy CP-even Higgs boson, which predominantly mediates annihilation and scattering, must have a coupling to weak gauge bosons at or below percent level to suppress the continuum gamma-ray spectrum below the limit from the Fermi-LAT data and the anti-proton spectrum constrained by the PAMELA data.  Discovering or excluding this CP-even Higgs boson at the LHC with a mass between 265 and 280 GeV and an enhanced diphoton branching ratio is crucial to test this scenario. 
\end{abstract}
\maketitle

\noindent
{\it{\textbf{Introduction.}}}
The way in which dark matter interacts with the Standard Model (SM) remains a mystery. Scenarios in which the photon or $Z$ boson are mediators with electroweak interaction strengths have already been excluded by dark matter direct detection experiments. The remaining natural mediator is the SM Higgs boson, resulting in the ``Higgs portal" dark matter scenario~\cite{Burgess:2000yq,McDonald:1993ex,Patt:2006fw}. With the recent discovery of a Higgs-like particle with a  mass of 125-126 GeV at the LHC~\cite{ATLAS:2012gk,CMS:2012gu}, the SM Higgs portal dark matter scenario is excluded for dark matter masses below 1 TeV except for the resonant region in which the dark matter mass is close to one half of the Higgs mass~\cite{LopezHonorez:2012kv}. This suggests that extensions to the simplest Higgs portal are worthy of further consideration.

The recent LHC Higgs data, which includes a relatively large diphoton branching ratio, also hints more generally at new physics in the Higgs sector. A simple way to extend the SM Higgs sector is to consider models with two Higgs doublets (2HDM's)~\cite{Branco:2011iw,Bai:2012ex,Altmannshofer:2012ar}. In such scenarios, the mixture of the two CP-even neutral scalar fields and general couplings of two Higgs doublets to fermions can dramatically modify the lightest Higgs boson properties~\cite{Davidson:2005cw,Pich:2009sp,DiazCruz:2010yq}. The increase of the light Higgs diphoton branching ratio in the generalized 2HDM, which we have called the 2HDM-X~\cite{Bai:2012ex}, can be realized in two ways: reducing the total width or increasing the coupling to two photons from the charged Higgs contributions.

In addition, there has recently been a hint of a gamma-ray line around 130 GeV from the galactic center that was found from analyzing the Fermi Gamma-Ray Space Telescope (Fermi-LAT) data~\cite{Weniger:2012tx,Bringmann:2012vr,Su:2012ft}. This hint has been confirmed by the Fermi-LAT collaboration with a smaller statistical significance. Using the reprocessed data, the peak has shifted to a slightly higher mass at $\sim$ 135~GeV~\cite{Fermi:talk}. Many dark matter models with additional charged particles have been constructed to explain this tentative Fermi-LAT 135 GeV gamma-ray line (see~\cite{Bringmann:2012ez} and references therein). 

In this letter, we propose a ``2HDM portal" dark matter scenario that can explain both the LHC Higgs data and the Fermi-LAT 135 GeV gamma-ray line. The characteristic feature of the scenario is that the heavy CP-even Higgs boson of the 2HDM-X model is the mediator for the SM particles interacting with the dark matter particles.  For a wide range of parameter space, the dark matter particles primarily annihilate into two gamma-ray lines with suppressed cross sections into the continuous gamma-ray and anti-proton spectra~\cite{Cohen:2012me,Buchmuller:2012rc,Mazziotta:2012ux}. 

Motivated by the possibility of realizing the first order electroweak phase transition for baryogenesis~\cite{Profumo:2007wc,Barger:2011vm}, we consider the dark matter to be a real SM singlet scalar field in this work, and defer the consideration of the case of fermionic dark matter for a future study.  For the case of SM singlet scalar dark matter, the interactions of the dark matter can be described by a renormalizable scalar potential with only a few new parameters.  We will see that for this scenario to accommodate both the collider and astrophysical data, the heavy neutral Higgs boson must have essentially a vanishing vector boson coupling, though with a strict mass range between 265-280 GeV.
\\

\noindent
{\it{\textbf{The 2HDM portal.}}}
We consider the two Higgs doublet model of~\cite{Bai:2012ex} (the 2HDM-X), which includes two complex scalar doublets of opposite hypercharge:
\beqa
\renewcommand{\arraystretch}{1.5}
{\bf \Phi}_1 &=& 
\left[\begin{array}{cc}
(v_1 + \phi_1^r + i \, \phi_1^i)/\sqrt{2}\,, &
\Phi_1^-
\end{array} \right]  \nn \\
 {\bf \Phi}_2 &=&
\left[\begin{array}{cc}
\Phi_2^+\,, &
(v_2 + \phi_2^r + i \, \phi_2^i)/\sqrt{2}
\end{array} \right].
\eeqa
As usual, $v_{\rm EW}^2 = v^2_1 + v^2_2 = (246~\mbox{GeV})^2$, and the ratio of the two Higgs vacuum expectation values is $\tan{\beta} \equiv v_2/v_1$.  

To extend the 2HDM-X to include dark matter, we include a real scalar field, $S$, which is assumed to obey a ${\mathbb Z}_2$ symmetry ($S\rightarrow -S$).\footnote{One can also consider a complex scalar, where the CP-odd component can be an equally viable dark matter candidate~\cite{Barger:2008jx}.  However, much of the phenomenological discussion can be mapped to the real scalar case, which we focus on here.} In the absence of CP violation, the scalar potential is given by
\be
V = V_\Phi  + V_S + V_{\Phi S},
\ee
where $V_\Phi$ is the CP-conserving 2HDM scalar potential,
\bea
V_\Phi &=&  m_1^2 \Phi_1^\dagger \Phi_1+m_2^2 \Phi_2^\dagger \Phi_2 - \left(m_{12}^2 \Phi_1^\dagger \tilde \Phi_2 + {\rm h.c.}\right) \nn\\ &+& {\lambda_1 \over 2} |\Phi_1^\dagger \Phi_1|^2+{\lambda_2 \over 2} |\Phi_2^\dagger \Phi_2|^2+ \lambda_3  |\Phi_1^\dagger \Phi_1 \Phi_2^\dagger \Phi_2|\\ &+&\lambda_4 |\Phi_1^\dagger \tilde\Phi_2 \Phi_2^\dagger \tilde\Phi_1| + {\lambda_5 \over 2} \left[(\Phi_1^\dagger \tilde\Phi_2)^2 + (\Phi_2^\dagger \tilde\Phi_1)^2\right],\nn
\eea
($\tilde{{\bf \Phi}}_{1,2}\equiv -{\bf \Phi}_{1,2}^* \,i\sigma_2 $), $V_S$ is given by 
\bea
V_S &=&  {1\over 2}m_S^2 S^2+{\lambda_S\over 4} S^4,
\eea
and $V_{\Phi S}$, which includes the interactions between the scalar singlet and the Higgs doublets, takes the form
\be
V_{\Phi S} =  {\delta_1\over 2} S^2 \Phi_1^\dagger \Phi_1+ {\delta_2\over 2} S^2 \Phi_2^\dagger \Phi_2 + {\delta_3\over 2} S^2(\Phi_1^\dagger \tilde \Phi_2 + {\rm h.c.})\,,
\ee
in which $\delta_3$ is taken to be real.  As we will see, the couplings $\delta_{1,2,3}$ are responsible for many observable effects.

The Yukawa interactions of the Higgs doublets with the SM fermions are assumed to take the following restricted form, as discussed in  \cite{Bai:2012ex}:
\beqa
- {\cal L} &=& y_u\,\overline{u}_R \, (\cos{\gamma_u}\,{\bf \Phi}_2 \,-\, \sin{\gamma_u}\,\tilde{{\bf \Phi}}_1)\,Q_L    \nonumber \\
&&\hspace{-3mm}+\,y_d\, \overline{d}_R\,(\cos{\gamma_d}\,{\bf \Phi}_1 \,+\, \sin{\gamma_d}\,\tilde{{\bf \Phi}}_2) \, Q_L   \nonumber \\
 &&\hspace{-3mm}+\, y_\ell\, \overline{e}_R\,{\bf \Phi}_1 \, L_L \,+\, \mbox{h.c.}\,,
\eeqa
in which the $y_{u, d, \ell}$ are $3\times 3$ matrices in family space, and the $\gamma_{u,d}$ lie within the range $0\leq \gamma_{u,d}\leq \pi$.  

At leading order, the CP-even $h_i \gamma\gamma$ effective couplings are given by
\bea
{\cal A}_i &=& {\alpha\,m_{h_i}^2\over 4 \pi v_{\rm EW}}\left[ \sum_{j=q,\ell,W^\pm} N_{c j} Q_j^2 \kappa_{ij} F_{j}(\tau_{ij})\right. \nn\\ 
&+& \left.{g_{h_i H^+H^-} v_{\rm EW}\over 2 m_{H^\pm}^2} F_0(\tau_{i,H^\pm}) \right],
\eea
where in the sum over each loop particle $j$, $N_{cj}$ is the color factor, $Q_j$ is the charge, $\kappa_{ij}$ is the coupling to $h_i$ relative to the SM coupling, the $F_j$ are the respective loop functions, and $\tau_{ij}={4 m_j^2/m_{h_i}^2}$.  The gluon amplitude follows similarly, but retains only the quark loop~\cite{Gunion:1989we}.
\begin{table}[htdp]
\renewcommand{\arraystretch}{1.6}
\caption{The tree level Yukawa couplings of the CP-even ($h,H$) and CP-odd ($A$) states with respect to the SM Higgs.}
\begin{center}
\begin{tabular}{c|cccc}
\hline \hline
 & $VV$ & $t\bar t$ & $b\bar b$ & $\tau^+ \tau^-$\\
\hline
$h$ & $\sin(\beta-\alpha)$ & $\cos(\alpha+\gamma_u)\over \sin(\beta+\gamma_u)
$ & $ - {\tan(\alpha-\gamma_d)\over \cos(\beta-\gamma_d)}
$ & $-{\sin \alpha\over \cos \beta} $\\
$H$ & $\cos(\beta-\alpha) $&  ${\sin(\alpha+\gamma_u)\over \sin(\beta+\gamma_u)}$ & $ {\cos(\alpha-\gamma_d)\over \cos(\beta-\gamma_d)}$ & ${\cos \alpha\over \cos \beta} $\\
$A$ & 0&  $i\gamma_5 A_u$ & $ i\gamma_5  A_d$ & $i \gamma_5 A_\ell $\\
\hline \hline
\end{tabular}
\end{center}
\label{tab:coup}
\end{table}%

The tree-level deviations of the Yukawa couplings of the neutral Higgs bosons from the SM Higgs boson are given in Table~\ref{tab:coup} and will generally referred to as $\kappa_i$, with mass index $i$.  The CP-odd and charged Higgs couplings scale with $A_u = \cot(\beta+\gamma_u)$, $A_d = \tan(\beta-\gamma_d)$ and $A_\ell = \tan\beta$.  The charged Higgs couplings are given by
\bea
g_{H^- t \bar b} &=& {\sqrt 2\over v}\left( m_t A_u P_L + m_b A_d P_R\right) \,, \\
g_{H^- \nu \tau^+} &=& {\sqrt 2\over v} m_\tau A_\ell P_R .
\eea
We exchange the dimensionless scalar potential couplings for the masses ($M_h, M_H, M_A, M_{H^\pm}, M_S$) and mixings (the usual CP-even Higgs mixing angle $\alpha$ and $\tan \beta$) while maintaining perturbativity of the scalar couplings.  \\

\begin{figure}[htbp]
\begin{center}
\includegraphics[scale=0.4, angle=0]{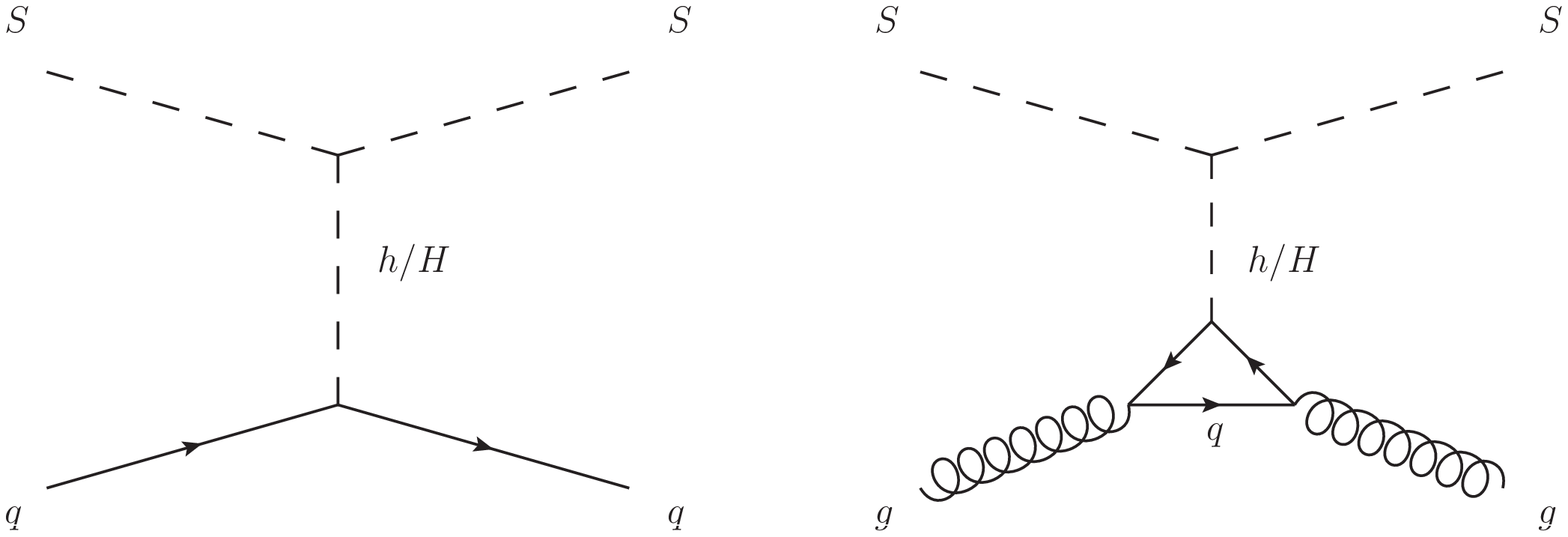}
\caption{Feynman diagrams for the scattering of dark matter off the partons of the proton.}
\label{fig:fdscatt}
\end{center}
\end{figure}

\noindent
{\it{\textbf{Dark Matter Observables.}}}
The direct detection of dark matter is observed via scattering off nuclei, which proceeds in this model through an exchange of neutral Higgs bosons, as shown in Fig.~\ref{fig:fdscatt}.  
The proton-$S$ spin-independent scattering rate is given by
\bea
\sigma^{SI}_{p-S} &=& {m_p^4\over 2\pi v_{\rm EW}^2(m_p+m_S)^2} \left\{ \sum_i {g_{SSh_i}\over m_{h_i}^2} [f_{pu} \kappa_i(u)\right. \nonumber \\
&+&\left. f_{pd} \kappa_i(d)+ f_{ps} \kappa_i(d)+{2\over 9}  f_{g} \kappa_i(g) ] \right\}^2,
\eea
in which the index $i$ accounts for both CP-even Higgs states.  The $SSh_i$ couplings, which arise from $V_{\Phi S}$, take the following values:
\bea
\frac{g_{SSh}}{v_{\rm EW}} \!=  \!\delta_2 \cos\alpha\sin\beta-\delta_1 \sin\alpha \cos \beta + \delta_3 \cos(\alpha+\beta)\nn \,,\\
\frac{g_{SSH}}{v_{\rm EW}} \!=  \!\delta_1 \cos\alpha\cos\beta+\delta_2 \sin\alpha \sin \beta+ \delta_3 \sin(\alpha+\beta).
\eea
Indirect detection begins with dark matter annihilation.  Dark matter in the present epoch is non-relativistic, and hence we treat annihilation as occurring in the static limit $v\to 0$.  Since here the dark matter is a scalar field that couples only to the Higgs boson, one can write a portion of dark matter annihilation rate in terms of the Higgs decay rate, as follows:
\bea
&&\sigma_{SS\to X \overline X} ~ v = {\Gamma_{h^{SM} \to X \overline X}(m_{h^{SM}}=2m_S)\over 2 m_S}\times\\ 
&&  \left| {g_{SSh}\, \kappa_1(X\overline X)\over 4m_S^2-m_h^2+i m_h \Gamma_h}+{g_{SSH} \, \kappa_2(X\overline X)\over 4m_S^2-m_H^2+i m_H \Gamma_H}\right|^2,  \nn
\eea
in which $X$ denotes any SM state that is not a Higgs boson or photon.  
\begin{figure}[t]
\begin{center}
\includegraphics[scale=0.4, angle=0]{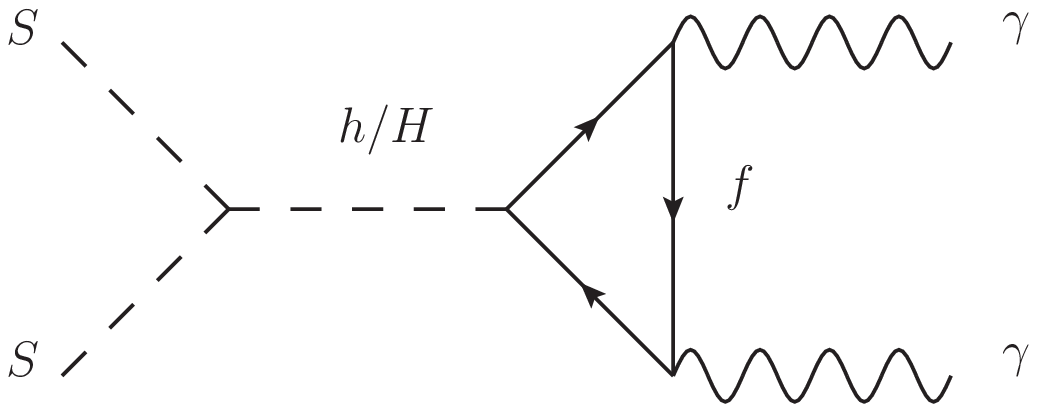}
\includegraphics[scale=0.4, angle=0]{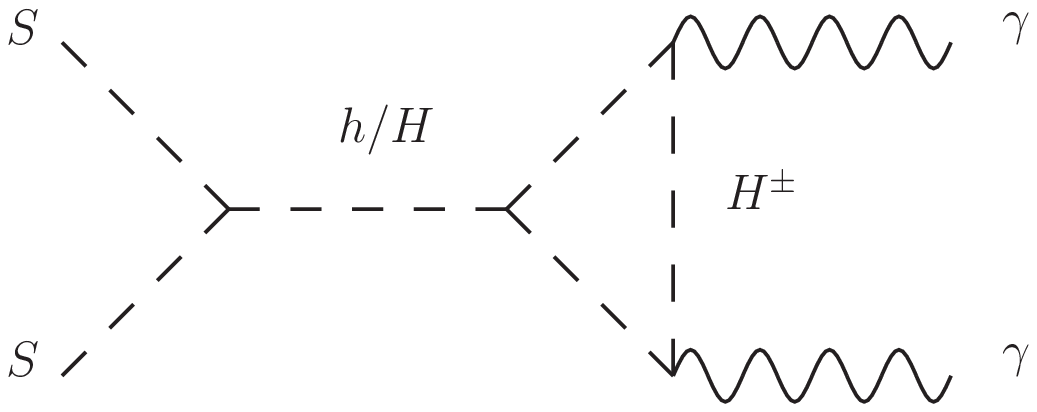}
\includegraphics[scale=0.4, angle=0]{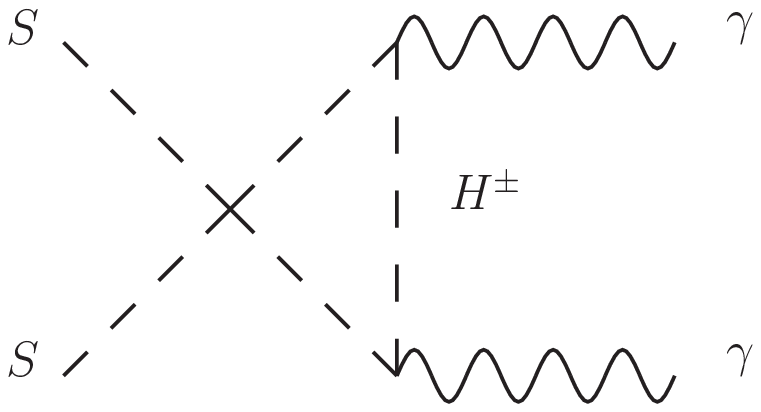}
\caption{Feynman diagrams for the annihilation of dark matter to photon pairs.}
\label{fig:fdaa}
\end{center}
\end{figure}
For photons, the annihilation rate is
\bea
&&\sigma_{SS\to \gamma\gamma} ~ v = 7.7\times10^{-8}  \left| {g_{SSh} \,\kappa_1(\gamma\gamma)\over 4m_S^2-m_h^2+i m_h \Gamma_h}\right.\\ 
&&\left.+{g_{SSH}\, \kappa_2(\gamma\gamma)\over 4m_S^2-m_H^2+i m_H \Gamma_H} + g_{SSH^+H^-} F_0\left({m_{H^\pm}^2\over m_S^2}\right) \right|^2,  \nn
\eea
which contains the processes shown in Fig.~\ref{fig:fdaa}.  The remaining annihilation modes include $SS\to hh$ and $hH, HH, AA, H^+H^-$ when kinematically accessible.    The Higgs bosons may acquire an additional decay width of
\be
\Gamma(h_i \to S S) = {g_{SSh_i}^2\over 32 \pi m_{h_i}} \sqrt{1-{4 m_S^2\over m_{h_i}^2}}\,,
\ee
which may be seen at colliders as missing energy. Indeed, searches for invisibly decaying Higgs bosons may be possible with early LHC data~\cite{Bai:2011wz}. \\

\noindent
{\it{\textbf{Fit to available data.}}}
We fit the 2HDM portal scenario outlined above to the available data using Bayesian inference, specifically adopting the Metropolis-Hastings Markov Chain Monte-Carlo algorithm; details for our implementation of this algorithm are given in~\cite{Barger:2008qd}.  Among its advantages include an efficient exploration of the model space as well as readily available posterior probability distributions. In addition to applying the experimental constraints listed below in our fit, we require all couplings remain perturbative: $g,\delta,\lambda < \sqrt{4\pi}$.

\noindent
$\bullet$ {\it Collider Data.}
We perform a Bayesian fit to the available collider data following the same method as~\cite{Low:2012rj}.  We require that the lightest CP-even mass eigenstate is the boson observed at the LHC and the Tevatron.  A simple combination of inclusive and selected exclusive channels provides the following distilled measurements~\cite{ATLAS:2012gk,CMS:2012gu,Tevatron:2012cn,ATLAS_Dec,CMS_Nov}:
\bea
\mu_{pp}(\gamma\gamma) &=& 1.7^{+0.3}_{-0.3}\,,\\
\mu_{pp}(VV) &=& 0.88^{+0.16}_{-0.16}\,,\\
\mu_{Vh}(b\bar b) &=& 1.1^{+0.4}_{-0.4}\,, \\
\mu_{gg}(\tau^+\tau^-) &=&1.1^{+0.8}_{-0.8}\,,\\
\mu_{VV}(\tau^+\tau^-) &=&0.58^{+0.74}_{-0.77}\,.
\eea
We also require the second CP-even Higgs state to fall below the present exclusion limits from ATLAS, which strongly disfavor a large vector boson coupling for moderate Higgs masses.  In addition, we include limits from the flavor changing decays $B_0\to X_s+\gamma+X$ at next to leading order~\cite{Ciuchini:1997xe} for consistency with the experimental measurement of $\mbox{BF}(B_0\to X_s+\gamma+X) = (3.55\pm0.26)\times 10^{-4}$~\cite{Amhis:2012bh}.  We further require consistency of the electroweak oblique parameters with $S=0.04\pm 0.09$, $T = 0.07 \pm 0.08$ with an 88\% positive correlation~\cite{Beringer:1900zz}.  Recent measurements of ${\rm BF}(B_s\to \mu^+\mu^-)$  have little impact on our scan as the new physics effects are suppressed for low $\tan\beta$.

\noindent
$\bullet$ {\it Dark Matter Data.}
We assume the Fermi $\gamma\gamma$ line at $E_\gamma=135$ GeV is astrophysical in origin and arises from dark matter annihilation with $m_S=135$ GeV.  We adopt the fitted value~\cite{Weniger:2012tx}
\be
\sigma_{\gamma\gamma} v = 2.27^{+0.65}_{-0.76}\, (1.27^{+0.37}_{-0.43} )\times 10^{-27}  \,{\rm cm}^3 {\rm s}^{-1},
\label{eq:fermiX}
\ee
which assumes an NFW (Einasto) dark matter galactic halo profile.  The astrophysical uncertainties for the required dark matter annihilation cross sections are fairly large. While we include the NFW value in our fit, we show the Einasto fit in our results as a comparison.  Generally, since the Einasto profile yields a lower fitted cross section, it can be more easily accommodated within this model. The Fermi collaboration analysis finds a similar line, but with lower significance.   In addition to the $SS\to\gamma\gamma$ line, a generic model will have annihilation to $SS\to Z\gamma$, where the $\gamma$ line is shifted to lower energy.  In this scenario, $E_\gamma=120$ GeV.  We impose a maximal cross section via Fermi data for this mode of~\cite{Fermi:talk}
\be
\sigma_{Z\gamma} v \lesssim 1.4 \times 10^{-27} \,{\rm cm}^3 {\rm s}^{-1}.
\ee
The Fermi satellite further includes limits on the secondary photons from possible dark matter annihilation in dwarf galaxies and the galactic halo~\cite{Mazziotta:2012ux}.  Furthermore, stringent limits on annihilation to hadronic final states can be placed from PAMELA $\bar p$ data.  For the mass we assume, the combined limits are strongest for annihilation to $b\bar b$ and $W^+W^-$.  We require the annihilation cross section for the most constrained modes to fall below~\cite{Asano:2012zv}
\bea
\label{eq:cont}
\sigma_{b\bar b} v &\lesssim& 4.2\times 10^{-26}\,  \rm{cm}^3 \rm{s}^{-1},\nn\\
\sigma_{W^+W^-} v &\lesssim& 3.8\times 10^{-26} \,\rm{cm}^3 \rm{s}^{-1},\\
\sigma_{\tau^+\tau^-} v &\lesssim& 1.4\times 10^{-25} \,\rm{cm}^3 \rm{s}^{-1}.\nn
\eea
Due to the dark matter mass indicated by the Fermi-LAT line, another important annihilation mode includes $SS\to hh$ annihilation which can proceed via $s$-channel $h/H$, $t$-channel $S$ and four-point $SShh$ processes.  This mode is taken into account by scaling the energy from dark matter annihilation by $1/2$ to account for halving the available phase space for the annihilation products, while the cross section limit increases by a factor weighted by the respective branching fraction for the final state particles.  For instance, for $b\bar b$, we have
\be
\sigma_{hh}(m_S)\cdot v = \sigma_{b\bar b}\left({m_S\over 2}\right) \cdot v~{1\over 2\,{\rm BF}(h\to b\bar b)} \,,
\ee
where either or both $h$'s can decay to $b\bar b$. We also apply the most restrictive direct detection measurement to date for  dark matter of moderate mass, which comes from the XENON experiment.  This gives a limit of $\sigma^{\rm SI} \lesssim 4\times 10^{-45} \,{\rm cm}^2$ for $m_S=135$ GeV at 95\% C.L.~\cite{Aprile:2012nq}.\\

\noindent
{\it{\textbf{Discussion.}}}
The LHC and Tevatron Higgs data appear to prefer a $hVV$ coupling consistent with custodial symmetry, since both the ratio in the $WW/ZZ$ and overall rates agree within measurement uncertainty with the SM.  These constraints, coupled with the null results for heavier Higgs masses, leave little room for a large $HVV$ coupling.   Therefore,  the key observation is that if  annihilation is mediated primarily through $H$, the  $H\to W^+W^-, ZZ$ modes are suppressed, allowing for an escape from the limits in Eq.~(\ref{eq:cont}).  Such a dominant annihilation through $H$ occurs when the $SSh$ coupling is suppressed, which happens when
\bea
\delta_3\! \!&&\approx {\delta_1 \sin\alpha\cos\beta-\delta_2\cos\alpha\sin\beta\over \cos (\alpha+\beta)} \\
  &&\approx  -{1\over 2}\left(\delta_1 \cot\beta+\delta_2\tan\beta\right)
 -{\Delta_V\over 4}\left({\delta_2\over \cos^2\beta}- {\delta_1\over  \sin^2\beta}\right),\nn
\eea
where $\Delta_V$, which is defined by $g_{hVV}^2 = 1-\Delta_V^2$, parameterizes the level of decoupling in the heavy Higgs sector.  

We also note that a suppression in the $HVV$ coupling from the SM allows a natural enhancement of the $SS\to \gamma\gamma$ line relative to $W^+ W^-$.  Indeed, the $\gamma\gamma$ line fit and continuum $\gamma$ ray constraints require at least
\be
{{\rm BF}(\phi\to \gamma\gamma)\over {\rm BF}(\phi\to W^+ W^-)} \gtrsim {\cal O}(10^{-2})\,,
\ee
for $m_\phi = 2 m_S = 270$ GeV.  In contrast, a SM Higgs mediating dark matter annihilation gives only
\be
{{\rm BF}(\phi\to \gamma\gamma)\over {\rm BF}(\phi\to W^+ W^-)} \sim {\cal O}(10^{-5})\,.
\ee
This enhancement thus does not occur in a simpler Higgs portal model with only one Higgs boson.

The $SS\to hh$ annihilation mode is kinematically accessible and is of importance due to the $h$ decay to secondary $\gamma$ and $\bar p$.\footnote{We enforce similar restrictions to the $SS\to H^+H^-$ and $Hh$ modes, when accessible.  The light charged Higgs is preferred to give an enhanced $h\to \gamma\gamma$ rate seen at the LHC.  The $AA$ mode can easily be closed by increasing $m_A$ above threshold.}  The constraints on this mode require a suppression in both $g_{SShh}$ and $g_{SSh}$ couplings, which take similar form $g_{SSh} \sim v~ g_{SShh}$.  Therefore, the suppression required for the $SS\to b\bar b$ and $W^+W^-$ channels also partially suppresses the $SS\to hh$ amplitudes involving the $s$-channel $h$ and four-point $SShh$ diagrams.  The remaining $s$-channel $H$ diagram is suppressed in the decoupling limit by the coupling
\be
g_{hhH} = \Delta_V\,  {{8 m_{12}^2/\sin(2\beta)}-2m_h^2-m_H^2\over v_{\rm EW}}\,.
\ee

\begin{figure}[htbp]
\begin{center}
\includegraphics[scale=0.5, angle=0]{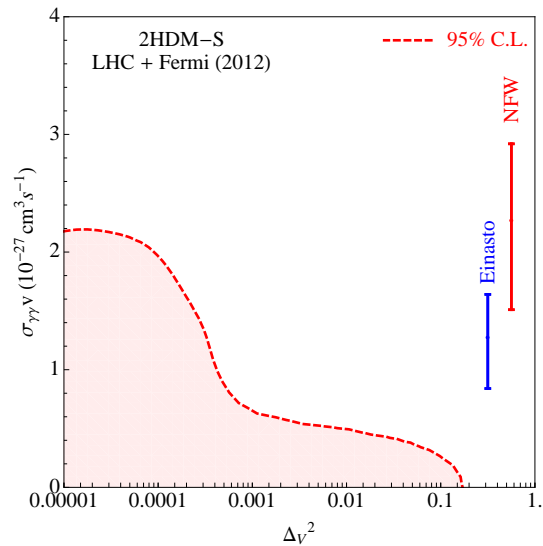}
\caption{The correlation of the decoupling limit parameter $\Delta_V^2$ with $\sigma_{\gamma\gamma} v$, which shows the importance of the suppressed $WWH$ and $ZZH$ couplings in fitting the Fermi-LAT data. The vertical lines illustrate the $1\sigma$ range for the Fermi-LAT line. }
\label{fig:decoup}
\end{center}
\end{figure}
The decoupling limit and the constraints from secondary $\gamma$ and $\bar p$ thus provide an explanation for the size of the Fermi-LAT $\gamma\gamma$ line within this scenario.  The importance of the decoupling limit is highlighted in Fig.~\ref{fig:decoup}, which shows that to reproduce the Fermi-LAT line, the decoupling limit must be realized to better than the mil-level.  

\begin{figure}[htbp]
\begin{center}
\includegraphics[scale=0.5, angle=0]{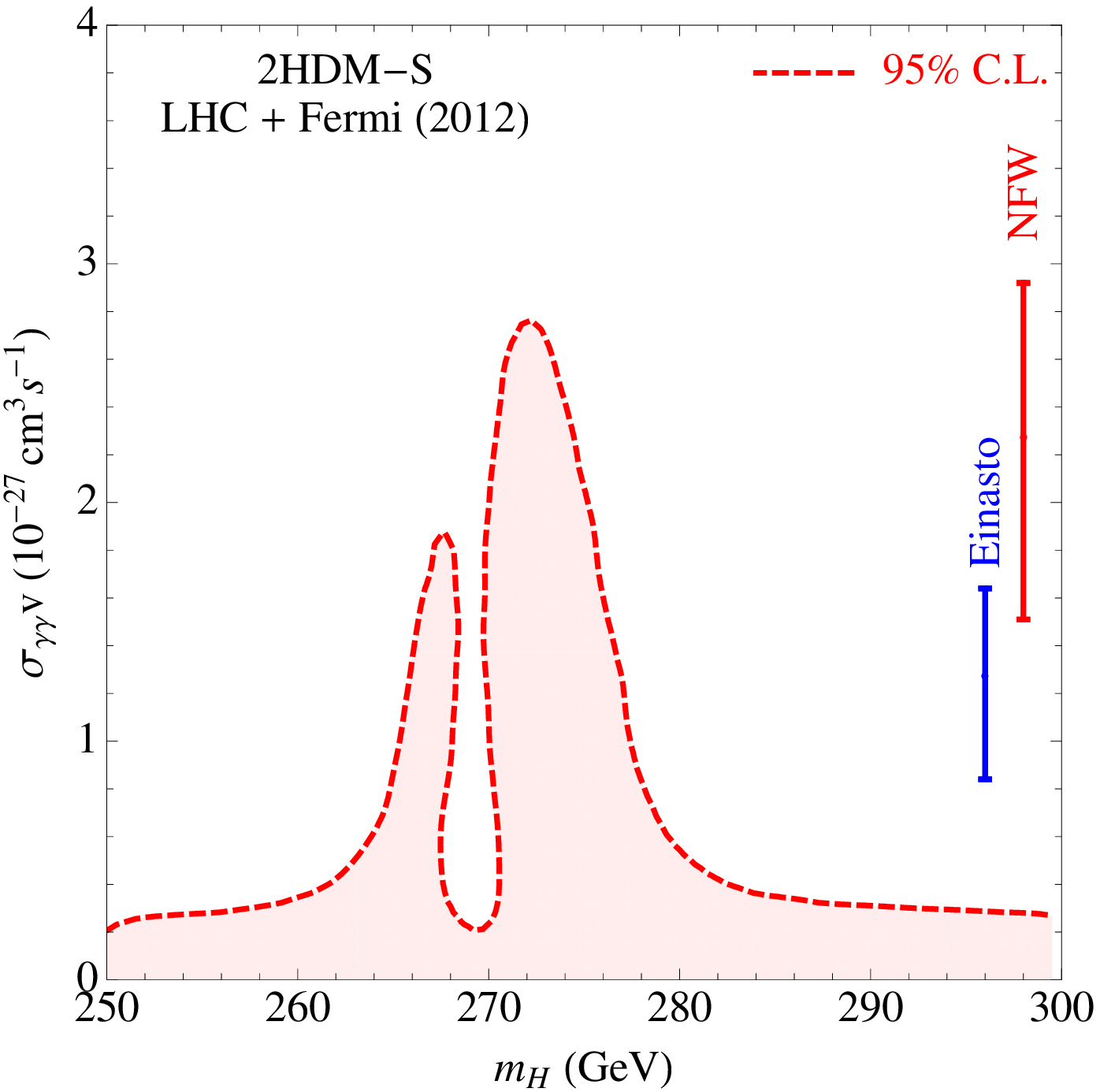}
\includegraphics[scale=0.5, angle=0]{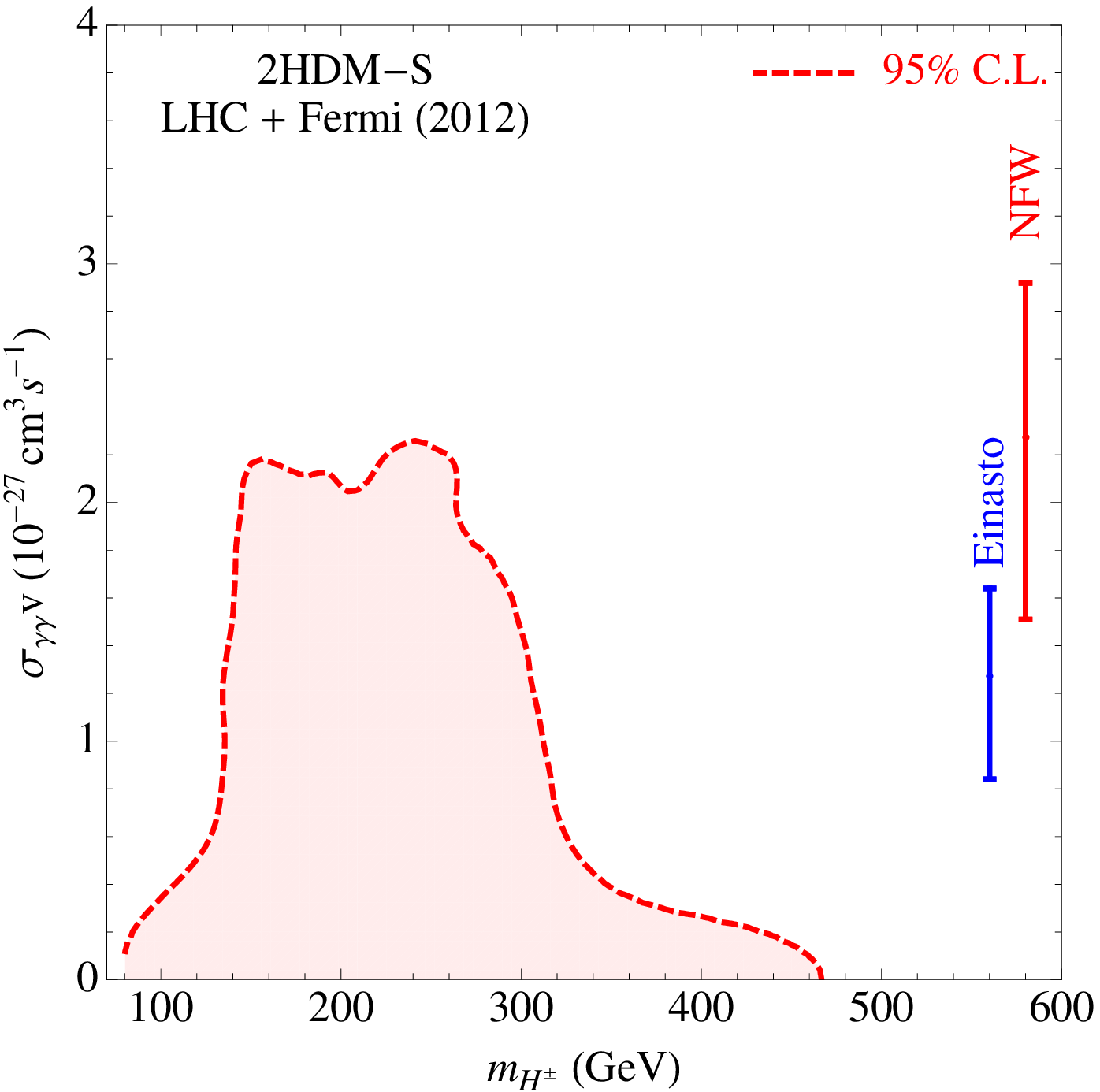}
\caption{The correlation of the heavy and charged Higgs masses with $\sigma_{\gamma\gamma} v$.   To explain the Fermi-LAT line, the heavy Higgs mass is likely to be near the $SS\to H$ resonance.  The vertical lines illustrate the $1\sigma$ range for the Fermi-LAT line. }
\label{fig:mH_sigaa}
\end{center}
\end{figure}
Due to the tight connection between the heavy Higgs boson and the associated annihilation cross sections, the Fermi-LAT line may also provide insight to the heavy Higgs mass.  In Fig.~\ref{fig:mH_sigaa}, we observe a tight region for $m_H$ that centers on the resonance $SS\to H$ from 265-280 GeV at  95\% C.L., but with a narrow gap at 270 GeV.  In this region, the $H$-resonant annihilation to $b\bar b$ increases beyond the Fermi limit.   The charged Higgs mass is expected to lie in a region of $140-320$ GeV at 95\% C.L., as required to elevate the $h\to \gamma\gamma$ and $SS\to \gamma\gamma$ rates.   We also note that if the $\gamma\gamma$ rate seen at the LHC subsides to the SM-like rate, the expected range for $m_H$ and $m_{H^\pm}$ does not appreciably change.\\

\noindent
{\it{\textbf{Summary.}}}
The 2HDM portal dark matter scenario, in which the 2HDM is extended with a real scalar singlet that plays the role of the dark matter, can explain both the LHC Higgs data and the Fermi-LAT gamma ray line data.  In this model, the heavy CP-even Higgs state is the mediator of the interactions of the dark matter with the SM fields.  The model predicts the following Higgs sector features: (i) 
the heavy Higgs sector is decoupled at the percent level or better, (ii) the heavy CP-even Higgs mass is preferentially between 265-280 GeV, and (iii) the charged Higgs mass is preferentially between 140 and 320 GeV.  
Furthermore, the production of secondary $\gamma$ and $\bar p$ are suppressed below present bounds, but it is not unrealistic to expect an observation of these modes soon as the $SS\to hh$ annihilation rate can be large.  

\vspace{3mm}

{\it{\textbf{Acknowledgements.}}}
VB, LE, and GS are supported by the U. S. Department of Energy under the contract DE-FG-02-95ER40896. YB is supported by startup funds from the UW-Madison. YB also thanks SLAC for its warm hospitality and the KITP, which
was supported in part by the National Science Foundation under Grant No. NSF PHY05-51164.

\end{document}